\documentclass[10pt, a4paper]{article}

\usepackage[final]{lrec-coling2024} 
\usepackage{algorithm,algcompatible,amsmath}
\usepackage{amsthm,amsopn,bm}
\usepackage[noend]{algpseudocode}
\usepackage{array}
\usepackage{balance}
\usepackage{booktabs}
\usepackage{bigstrut}
\usepackage{booktabs}
\usepackage{blindtext}
\usepackage{caption}
\usepackage{color,soul}
\usepackage{enumitem}

\usepackage{etoolbox}
\usepackage{footnote}
\usepackage{graphicx}
\usepackage{longtable}
\usepackage{multirow}
\usepackage{mathtools}
\usepackage{multibib}
\usepackage{natbib}
\usepackage{stmaryrd}
\usepackage{subcaption}
\usepackage{soul}
\usepackage{tabularx}
\usepackage{tabulary}
\usepackage{tikz}
\usepackage{titlesec}
\usepackage{url}
\usepackage{verbatim}
\usepackage{xcolor,colortbl}

\newcommand\mf[1]{\mathbf{#1}}

\newcommand\mr[1]{\mathrm{#1}}

\title{M3: A Multi-Task Mixed-Objective Learning Framework for Open-Domain Multi-Hop Dense Sentence Retrieval\\ \vspace*{.5\baselineskip}}

\name{
    Yang Bai,
    Anthony Colas,
    Christan Grant,
    Daisy Zhe Wang
    } 

\address{
        The University of Florida \\
        Gainesville, Florida, USA \\
         \{baiyang94, acolas1, christan, daisyw\}@ufl.edu\\}

\abstract{
In recent research, contrastive learning has proven to be a highly effective method for representation learning and is widely used for dense retrieval. However, we identify that relying solely on contrastive learning can lead to suboptimal retrieval performance. On the other hand, despite many retrieval datasets supporting various learning objectives beyond contrastive learning, combining them efficiently in multi-task learning scenarios can be challenging. In this paper, we introduce \textbf{M3}, an advanced recursive \textbf{M}ulti-hop dense sentence retrieval system built upon a novel \textbf{M}ulti-task \textbf{M}ixed-objective approach for dense text representation learning, addressing the aforementioned challenges. Our approach yields state-of-the-art performance on a large-scale open-domain fact verification benchmark dataset, FEVER. 
Code and data are available at: \url{https://github.com/TonyBY/M3}
\newline  }

\begin{document}

\maketitleabstract

\section{Introduction}
\label{sec:introduction}
\begin{figure}[t]\setlength{\textfloatsep}{10pt}
    \centering
    \includegraphics[width=\linewidth]{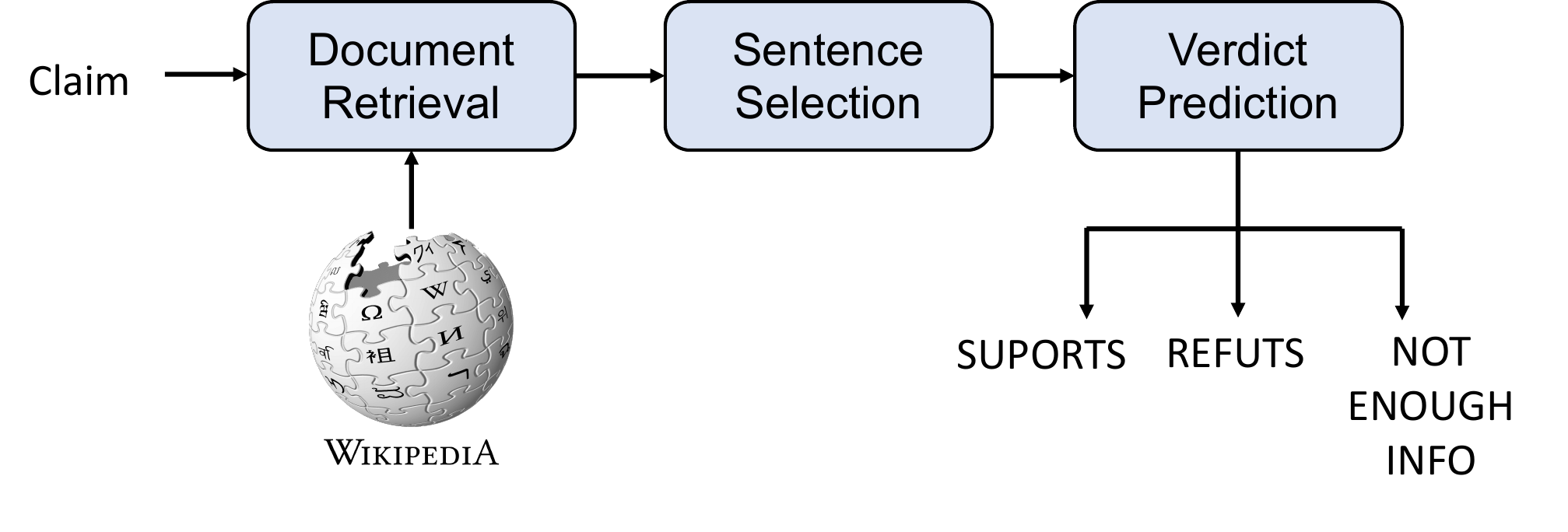}
    \caption{Canonical thee-stage fact verification framework.}
    \label{fig:three-stage}
\end{figure}

Open-domain fact verification~\citep{thorne-etal-2018-fever, Jiang2020HoVerAD, Bai2023MythQAQL} is a challenging task where single-hop or multi-hop sentence-level evidence for a given claim needs to be extracted from a large pool of documents to verify human-generated claims (See Figure~\ref{fig:FEVER-example} for an example from the FEVER dataset). A three-stage approach is commonly used to solve the problem (see Figure~\ref{fig:three-stage}) \citep{thorne-etal-2018-fever, yoneda-etal-2018-ucl, hanselowski-etal-2018-ukp, Nie2018CombiningFE, Zhong2019ReasoningOS, zhou-etal-2019-gear, Soleimani-BERT-FEVER, subramanian-lee-2020-hierarchical, Jiang2021ExploringLE, Krishna2021ProoFVerNL, Fajcik2022ClaimDissectorAI, dehaven-scott-2023-bevers}. In the first step, the retriever produces a list of $n$ candidate documents given a claim. From the top-n documents, a sentence reranker selects the top-k sentences. Lastly, a claim classifier predicts the claim verdict based on the top-n sentences. 

Retrieval models have traditionally relied on term-based information retrieval (IR) methods \citep{thorne-etal-2018-fever, yoneda-etal-2018-ucl, Nie2018CombiningFE, Jiang2021ExploringLE, dehaven-scott-2023-bevers}, which do not capture the semantics of a claim beyond lexical matching and remain a key bottleneck. In contrast, recent works train neural network-based encoders to obtain dense representations of queries and documents in vector spaces and then use maximum inner-product search (MIPS) to complete the retrieval \citep{Karpukhin2020DensePR, 10.1145/3397271.3401075, Qu2020RocketQAAO, ren-etal-2021-rocketqav2, xiong2021approximate, xiong2021answering, Wu2021SentenceawareCL, zhang2022AR2}. When compared to traditional IR approaches, these dense retrievers have demonstrated significant improvements.

\begin{figure}[t]
    \centering
    \scalebox{1.0}{
    \begin{tabular}{@{}p{21em}}
        \toprule 
        \textbf{Claim: \textcolor{blue}{Sheryl Lee} has yet to appear in a \textcolor{red}{\textit{film}} as of \textcolor{red}{\textit{2016}}.} \\
        \midrule
        \textbf{Evidence Documents}\\\\
        
        
        \textbf{Doc1: \textcolor{blue}{Sheryl Lee}}\\
        ~~~\textcolor{red}{\textit{In 2016}}, she appeared in \textcolor{brown}{\textbf{Caf\'e Society}}, and also completed the Showtime revival of Twin Peaks (2017), reprising her role of Laura Palmer.\\
        \\
        
        \textbf{Doc2: \textcolor{brown}{Caf\'e Society}}\\
        ~~~\textcolor{red}{\textit{Caf\'e Society is a}} 2016 American romantic comedy-drama \textcolor{red}{\textit{film}} written and directed by Woody Allen .\\
        \midrule
        \textbf{Verdict:} Refuted\\
        \bottomrule
    \end{tabular}}
    \caption{A FEVER example where multi-hop sentence-level evidence from multiple Wikipedia documents is required for verification. 
    }
    \label{fig:FEVER-example}
\end{figure}

There are, however, some issues with current dense information retrieval models. First of all, they are trained on datasets at the document/passage level. This can increase the possibility of learning suboptimal representations due to internal representation conflicts~\cite{Wu2021SentenceawareCL}. In particular, a passage can be organized by multiple semantically different sentences. It is not optimal to model a passage like this as a unified dense vector. Moreover, each document/passage consists of multiple sentences, from which multiple semantically distant queries (questions, claims, etc.) can be derived. In contrastive learning frameworks, that can benefit from large batch sizes when doing in-batch negative sampling, such a one-to-many problem can lead to severe conflicts when two conflicting claims with the same context document/passage are sampled within the same batch~\cite{Wu2021SentenceawareCL}. 
Hence, to avoid the above-mentioned conflicts, we propose to use dense sentence-level retrieval as the first-level retriever to replace the traditional document retrieval modules in the canonical open-domain fact verification pipeline.

Furthermore, we observe that current dense information retrieval models rely solely on contrastive objectives, which could prevent the models from learning better representations and subsequently result in suboptimal recall. 
On the other hand, despite many retrieval datasets supporting various learning objectives beyond contrastive learning, combining them efficiently in multi-task learning scenarios can be challenging. In this paper, we introduce \textbf{M3}, an advanced recursive \textbf{M}ulti-hop dense sentence retrieval system built upon a novel \textbf{M}ulti-task \textbf{M}ixed-objective approach for dense text representation learning, addressing the aforementioned challenges. Our approach yields state-of-the-art performance on a large-scale open-domain fact verification benchmark dataset, FEVER.

Contributions of this paper include:
\begin{itemize}
    \item {
    We present an advanced recursive multi-hop dense sentence retrieval system (M3) based on a novel dense sentence representation learning method, which achieves state-of-the-art multi-hop retrieval performance on the FEVER dataset.
     }
    \item{
    We propose a novel dense sentence representation learning method (M3-DSR) based on multi-task learning and mixed-objective learning frameworks that significantly outperforms strong baselines such as BM25~\cite{Yang2017AnseriniET} and DPR~\cite{Karpukhin2020DensePR} on sentence-level retrieval.
    }
    \item {
     We introduce an efficient heuristic hybrid ranking algorithm for combining retrieved single-hop and multi-hop sentence evidence, which shows substantial improvements over previous methods.
     }
    \item{
    We developed an end-to-end multi-hop fact verification system based on M3 that achieves state-of-the-art performance on the FEVER dataset.
    }
\end{itemize}

\section{Background and Related Works}
\label{sec:related-works}
\subsection{Dense Text Retrieval}
\noindent Dense Text Retrieval (DTR)~\citep{Karpukhin2020DensePR, 10.1145/3397271.3401075, xiong2021approximate, Wu2021SentenceawareCL, zhang2022AR2} has gained significant attention in recent years due to its potential to revolutionize sparse retrieval methods such as TF-IDF~\cite{Ramos2003UsingTT} and BM25~\citep{Robertson1994OkapiAT, Yang2017AnseriniET} on the document retrieval task. Contrastive learning, at its core~\citep{Hadsell2006DimensionalityRB, Reimers2019SentenceBERTSE, Karpukhin2020DensePR}, aims to learn effective representations by contrasting similar and dissimilar pairs of data. In the context of DTR, this translates to training models to distinguish between relevant and non-relevant document-query pairs.


In recent studies, \citet{akkalyoncu-yilmaz-etal-2019-cross} and \citet{Wu2021SentenceawareCL} propose to improve dense passage retrieval based on sentence-level evidence. In particular, \citet{Wu2021SentenceawareCL} investigated contrastive conflicts in the contrastive learning framework when performing document/passage-level representation learning as discussed in Section~\ref{sec:introduction}.

In contrast, we propose a simple approach that bypasses such conflicts by performing dense sentence-level retrieval in combination with multi-task, mixed-objective learning that shows stronger empirical performance. 

\begin{figure*}[tb]
    \centering
    \includegraphics[scale=0.47]{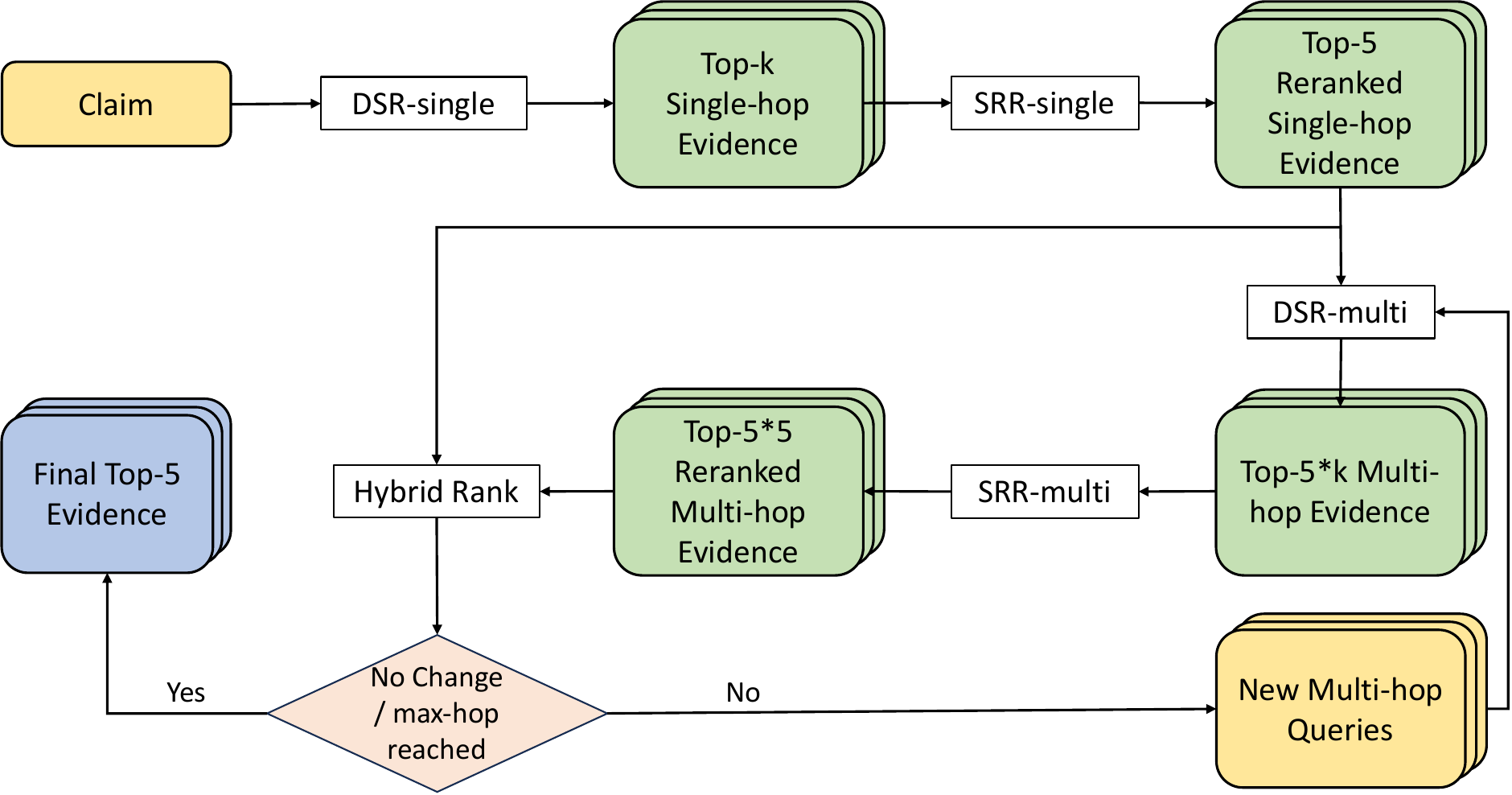}
    \vspace{5pt}
    \caption{M3 iterative dense sentence retrieval pipeline. DSR refers to the dense sentence retrieval model; SRR refers to the sentence reranking model; *-single and *-multi indicate whether the model is trained on single-hop or multi-hop examples. When no specific number of hops is given, the multi-hop retrieval process continues until the top-5 hybrid-ranked sentences stop changing.} 
    \label{fig:m3_pipeline}
\end{figure*}

\subsection{Multi-hop Text Retrieval}
\noindent The multi-hop text retrieval method is crucial to complex question-answering~\citep{Nie2019RevealingTI, xiong2021answering, 10096119} and complex fact-verification~\citep{thorne-etal-2018-fever, yoneda-etal-2018-ucl, hanselowski-etal-2018-ukp, Nie2018CombiningFE, Jiang2021ExploringLE, dehaven-scott-2023-bevers} tasks where evidence is aggregated from multiple documents before logical reasoning or multi-hop reasoning is applied to infer the answer or a verdict. In prevailing approaches~\citep{Nie2018CombiningFE, Asai2019LearningTR}, a document graph is constructed based on entity linking or hyperlinks found in the underlying Wikipedia corpus. These methods, however, might not be generalizable to new domains, where entity linking might perform poorly, or hyperlinks might be sparse\cite{xiong2021answering}.

With a recursive framework, MDR~\cite{xiong2021answering} applies dense retrieval to the multi-hop setting. Utilizing efficient MIPS methods, it iteratively encodes the question and previously retrieved documents as a query vector and retrieves the next relevant documents. \citet{aly-vlachos-2022-natural} propose a retrieve-and-rerank method, AdMIRaL, consisting of a retriever that jointly scores documents in the knowledge source and sentences from previously retrieved documents and achieves the state-of-the-art multi-hop document retrieval recall on the FEVER dataset. However, to ensure efficiency, AdMIRaL's first-stage retriever uses sparse retrieval (i.e., BM25), which may sacrifice retrieval recall.

In our work, we combine the advantages of MDR and AdMIRaL to develop M3 for dense multi-hop search using a recursive retrieve-and-rerank framework. We also propose a hybrid ranking algorithm to jointly rank the single-hop and multi-hop retrieval results and achieve better overall retrieval recall.

Moreover, unlike MDR and AdMIRaL, which focus only on the retrieval of multi-hop document-level evidence at the first stage, M3 is capable of achieving state-of-the-art retrieval performance for both sentence-level and document-level evidence, which is more challenging and offers more fine-grained evidence that is crucial for downstream multi-hop inference. Our fact-verification system based on M3 achieves the highest claim classification accuracy on the blind FEVER testing dataset.

\section{Method}\label{sec:Method}
\subsection{Overview}
\noindent Our work focuses on improving the retrieval component of open-domain fact verification. With a claim $c$ in natural language and a collection of $M$ text documents, the retrieval module needs to retrieve a ranked list of sentence-level evidence $S: \{s_1, s_2, ..., s_k\}(k << M)$ that provides sufficient information for downstream logical inference components to determine whether the claim $c$ is supported, refuted, or unverifiable by the facts in the corpus. It is important to note that $M$ can be very large (for example, in our setting, there are over 5 million documents with over 25 million sentences), and $k$ should be small ($k=5$ in the FEVER setting).

Our multi-hop dense sentence retriever M3 uses an iterative sentence-level retrieve-and-rerank scheme to recursively retrieve evidence (see Figure~\ref{fig:m3_pipeline}). The sentence retrieval probability at each step depends on the previous retrievals, i.e., $P(s_t|c,s_1, ..., s_{t-1})$. In practice, this probability is calculated as $P(s_t|q_{t-1})$ where $q_{t-1}=c\oplus s_1 \oplus...\oplus s_{t-1}$, and $\oplus$ refers to concatenation operator. 
When $t=1$, the retrieval probability is only conditioned on the original claim. 

M3 differs from the existing multi-hop dense document retrieval method\cite{xiong2021answering} in four ways: 1) finer retrieval granularity: document-level -> sentence-level, 2) we add reranking after each step of retrieval, and 3) when combining single-hop and multi-hop retrievals, a novel hybrid ranking algorithm is used, 4) we train a novel dual-encoder model using multi-task and mixed-objective learning to learn better dense text representations that yield higher retrieval recalls.

\begin{figure*}[t]
    \centering
    \includegraphics[scale=0.36]{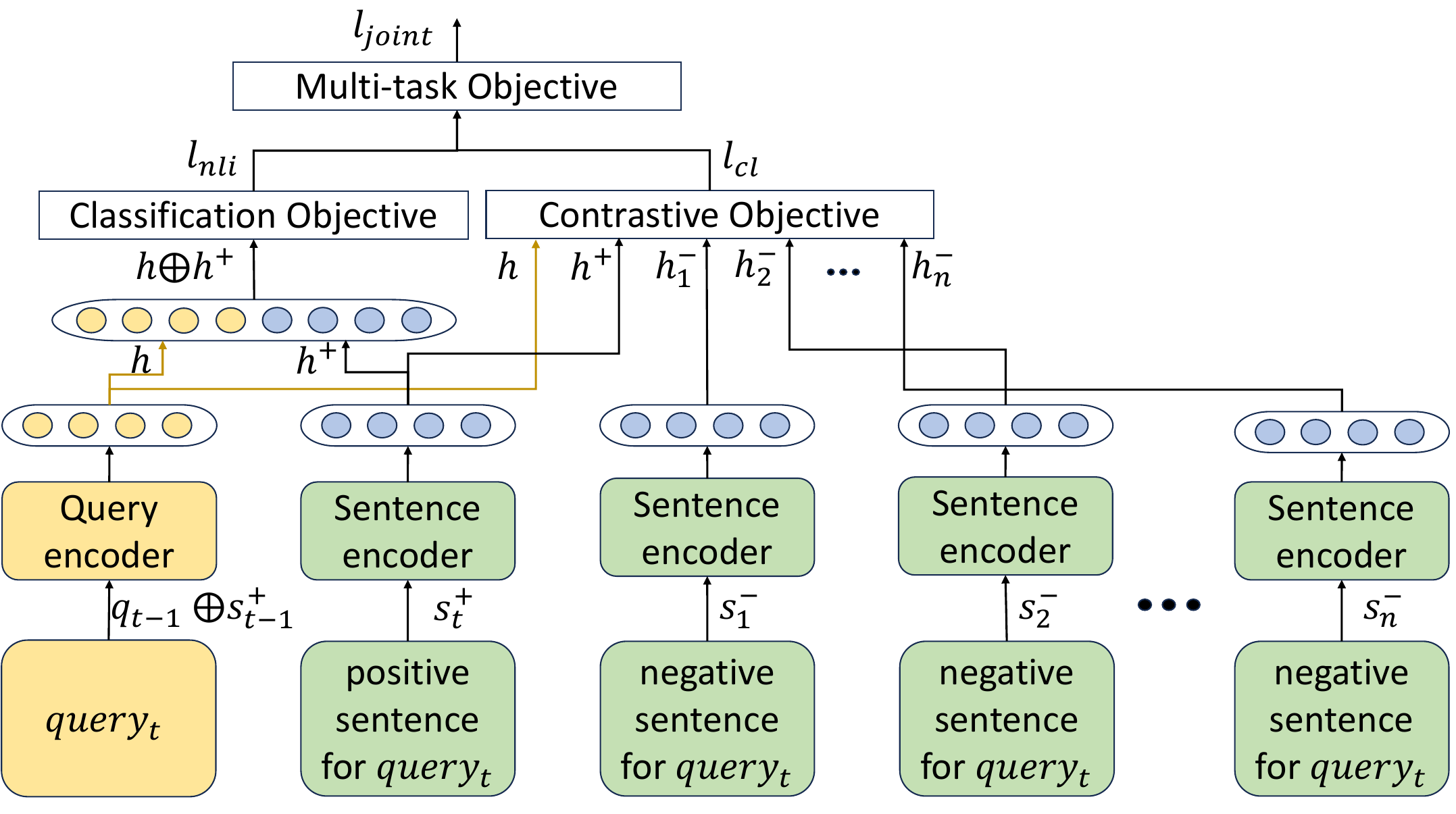}
    \caption{M3-DSR multi-task learning framework. When $t=1$ (i.e., first-hop), the input of the query encoder is the original claim $c$. 
    }
    \label{fig:m3_dsr}
\end{figure*}

\subsection{Dense Sentence Retrieval}
\noindent Dense sentence retrieval aims to learn low-dimensional and continuous representation for the queries and sentences in the corpus in order to efficiently retrieve the top-k sentence-level evidence through an approximate nearest neighbor (ANN) \cite{Johnson2017BillionScaleSS} method. 

We train our dense sentence retrievers (M3-DSR) with a novel multi-task mixed-objective learning method. Using this method, we are able to learn better sentence representations that yield better retrieval recall (see more ablation studies Section~\ref{sec: analysis}).

\vspace{-2pt}
\subsubsection{Multi-task Learning} 
\noindent In the FEVER dataset, evidence and verdict annotations are given for each claim. Naturally, we explore training better text encoders with the FEVER dataset through multi-task learning with two objectives, contrastive and (claim) classification. Figure~\ref{fig:m3_dsr} shows an overview of our multi-task learning framework.\\

\noindent\textbf{Contrastive Learning Objective} The contrastive objective is implemented as in~\citep{Karpukhin2020DensePR, gao-etal-2021-simcse}, where each input query $x_i$ is paired with a positive example $x^+_i$ and n negative examples $\{x^-_{i, 1}, x^-_{i, 2}, ..., x^-_{i, n}\}$. We also use the in-batch negative sampling~\cite{Karpukhin2020DensePR} taking other examples in the same batch as “negatives”, and the model predicts the positive one among negatives to approximate the softmax over all examples. Let $\mf{h}_i$ and $\mf{h}_i^+$ denote the  representations of $x_i$ and $x_i^+$, the training objective $\ell_i$ is then defined as:
\begin{equation}
    \label{eq:sup_objective}
    \begin{aligned}
        \ell_{cl\_i} = - \log \frac{e^{\mr{sim}(\mf{h}_i,\mf{h}_i^+ )/ \tau }}{\sum_{j=1}^N\left(e^{\mr{sim}(\mf{h}_i,\mf{h}_j^+)/\tau}+e^{\mr{sim}(\mf{h}_i,\mf{h}_j^-)/ \tau}\right)}.
    \end{aligned}
\end{equation}

\noindent where $\tau$ is a temperature hyperparameter, $N$ is batch size and $\mr{sim}(\mf{h}_1,\mf{h}_2)$ is the inner product $\mf{h}_1^\top \mf{h}_2$.
The input sentences are encoded by a transformer language model: 
$\mf{h} = f_{\theta}(x)$. Specifically, we obtain $\mf{h}$ for a special token $[CLS]$, which represents the whole input sequence.

 
When sampling negative examples, we follow \cite{Karpukhin2020DensePR} using BM25~\cite{10.1145/3404835.3463238} to retrieve top sentences from the whole corpus that are not included in the evidence annotation set. 
In addition, we filter these top negative examples with a more complex attention-based model to eliminate those that are too close to the claims to avoid including too many false negatives. Specifically, we use an off-the-shelf pre-trained sentence ranker that scores sentences based on their semantic similarity to the query. An empirical threshold is set based on extensive observation. We show that reducing the false negative examples in the training data is crucial to contrastive learning (see Section~\ref{sec: analysis} for more details). \\

\noindent\textbf{Classification Objective}
\noindent The encoded query $\mf{h_i}$ and positive example $\mf{h^+_i}$ is used to calculate the claim label probability $P(y|(\mf{h_i}, \mf{h^+_i}))$:
\begin{equation}
    \label{eq:verdict_prob}
    \begin{aligned}
        P(y|(\mf{h_i}, \mf{h^+_i})) = softmax_y(Linear(\mf{h_i} \oplus \mf{h^+_i})).
    \end{aligned}
\end{equation}

\noindent  where $\oplus$ refers to concatenation operator.

\noindent The claim classification (NLI) loss is then defined as:
\begin{equation}
    \label{eq:nli_loss}
    \begin{aligned}
        \ell_{nli\_i} = CrossEntropy(y^*, P(y|(\mf{h_i}, \mf{h^+_i}))).
    \end{aligned}
\end{equation}

\vspace{5pt}

\noindent\textbf{Multi-Task Objective}
\noindent Our multi-task objective is a linear combination of contrastive loss and classification loss:
\begin{equation}
    \label{eq:joint_loss}
    \begin{aligned}
        \ell_{joint\_i} = \alpha * \ell_{cl\_i} + \beta * \ell_{nli\_i}.
    \end{aligned}
\end{equation}
\noindent where hyperparameters: $\alpha$ and $\beta$ are $\in [0, 1]$

\subsubsection{Mixed-Objective Learning}
\begin{figure}[t]
    \centering
    \includegraphics[width=\linewidth]{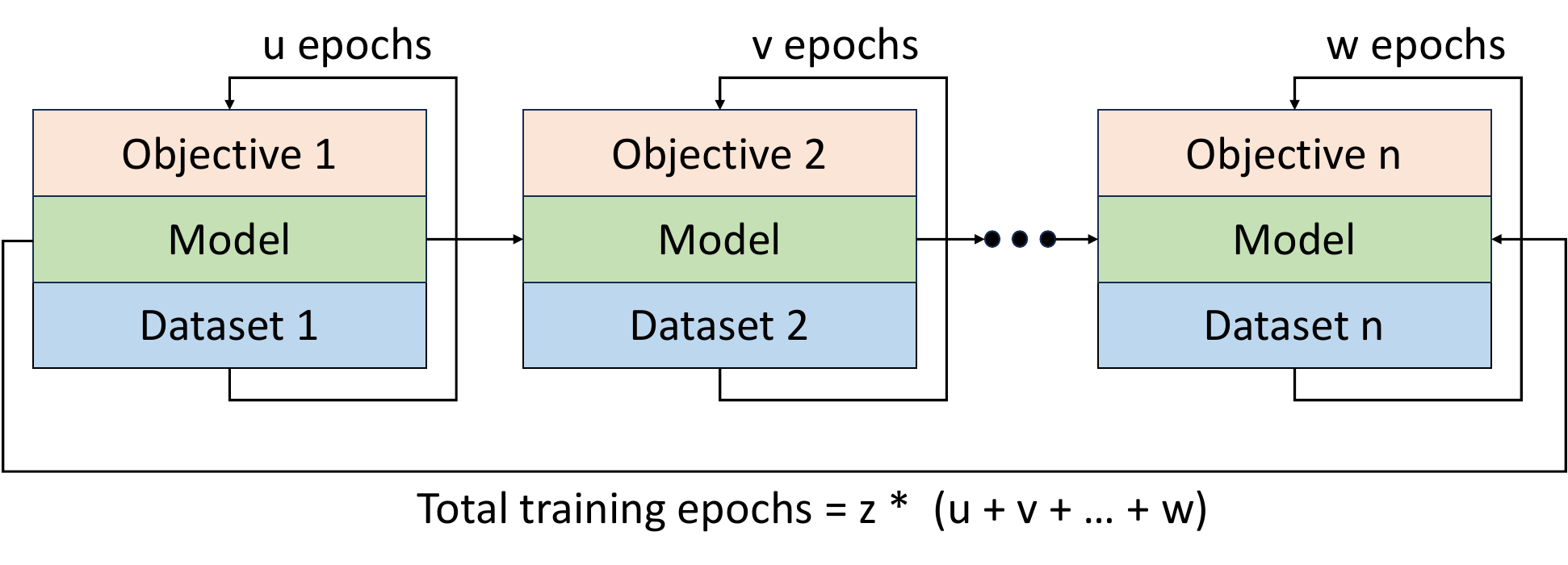}
    \caption{M3-DSR mixed-objective learning framework. The same model is trained with different dataset-objective combinations sequentially. 
    }
    \label{fig:m3_mix}
\end{figure}

\noindent Many datasets are proposed for training encoders for open-domain dense text retrieval~\citep{Kwiatkowski2019NaturalQA, Joshi2017TriviaQAAL, Berant2013SemanticPO, Baudis2015ModelingOT, Rajpurkar2016SQuAD1Q, thorne-etal-2018-fever}, however, they may support different objectives (e.g., contrastive, question answering, natural language inference, or multi-task). To make maximum use of these valuable datasets, we developed a framework that allows dense encoders to be trained over these datasets with different user-defined objectives and intervals/epochs (see Figure~\ref{fig:m3_mix} for more details). The framework is not trivial as it allows us to optimize the model across multiple datasets that support different objectives more flexibly and conveniently with access to hyperparameters such as the objective and appearance frequency of each dataset during training. In extensive experiments, this framework demonstrated its effectiveness in improving retrieval recall.

\subsection{Sentence Reranking}
\noindent In sentence reranking, the top retrieved sentences from the previous step are ranked again using a more sophisticated method.
Following~\cite{dehaven-scott-2023-bevers}, we reformulate this task as a sentence pair classification task: given a query-sentence pair $(q, s)$, predict labels from \{\texttt{SUPPORTS}, \texttt{REFUTES}, \texttt{NOT ENOUGH INFO (NEI)}\}. The relevancy score for sentence ranking is calculated as:
\begin{equation}
    \label{eq:srr_score}
    \begin{aligned}
        Score(s) = 1- softmax_{NEI}(
        h_{q \oplus s})
    \end{aligned}
\end{equation}
where $h_{q \oplus s}$ is the embedding of a concatenated query-sentence pair encoded by a language model. $sofmax_{NEI}$ calculates the normalized confidence score over the \texttt{NEI} (irrelevant) class.

From the top retrievals in the last step, we sample the negative (\texttt{NEI}) sentences to create the training data (see Section~\ref{sec: Experimental_Setup} for more details).

    \begin{algorithm}[tb]
        \begin{center}
            \footnotesize
    	\begin{algorithmic}[1]
    		\Require
                (1) $ScoreMap_{single}$, a dictionary that saves the top single-hop retrievals in key-value pairs, e.g., $\{\cdots, se_i: sc_i, \cdots\}$, where key ($se_i$) is a sentence id, and value ($sc_i$) is the corresponding score acquired by Equation~\ref{eq:srr_score};
        
                ~(2) $SequenceList$, a list of top multi-hop retrieval paths that consist of $t$ id-score pairs, each pair representing one step of iterative retrieval results of t-hops, e.g., 
                $[\cdots, \underbrace{((se_i, sc_i), (se_j, sc_j), \cdots, (se_k, sc_k))}_\text{t}, \cdots]$.
    
                ~~(3) $mth$ and $\gamma$ $\in (0, 1]$, hyperparameters that need to be tuned.
                
                
    		\Function{hybrid\_rank}{$ScoreMap_{single}$, 
      
                ~~~~~~~~~~~~~~~~~~~~~~~~~~~~$SequenceList$, $mth$, $\gamma$}
                \State
                $ScoreMap_{multi}=\{\}$
                \For{$seq$ in $SequenceList$}
                
                \State
                $seq\_score$ = \textit{Product}([$p[1]$ for $p$ in $seq$])
                \If{$seq\_score$ < $mth$}
                \State
                    continue
                \EndIf
                    \For{$p$ in $seq$}
                    \State
                    \textbf{if} $p[0]$ not in $ScoreMap_{multi}.keys()$ or 
                    
                    ~~~~~~~~~~~~$seq\_score$ > $ScoreMap_{multi}[p[0]]$
                    \State
                    \textbf{then}
                    \State
                    ~~~$ScoreMap_{multi}[pair[0]]$ = $seq\_score$
                    \EndFor
                \EndFor
                \State
                NormalizeScores($ScoreMap_{single}$)
                \State
                NormalizeScores($ScoreMap_{multi}$)
                \State
                $ScoreMap_{hybrid}=\{\}$
                \State
                \textbf{for}
                $id$ in union(set($ScoreMap_{single}.keys()$), 
               
                ~~~~~~~~~~~~~~~~~~~~~~~set($ScoreMap_{multi}.keys()$)
                \State
                \textbf{do}
                \State
                    ~~~~~~\textbf{if} $id$ not in $ScoreMap_{single}.keys()$ \textbf{then}
                    \State
                    ~~~~~~~~~~$ScoreMap_{single}[id]$ = 
                    
                    \hfill \textit{minValue}($ScoreMap_{single}$)
    
                \State
                    ~~~~~~\textbf{if} $id$ not in $ScoreMap_{multi}.keys()$ \textbf{then}
                    \State
                    ~~~~~~~~~~$ScoreMap_{multi}[id]$ = 
                    
                    \hfill \textit{minValue}($ScoreMap_{multi}$)
    
                \State
                ~~~~~~$ScoreMap_{hybrid}[id]$ = $ScoreMap_{single}[id]$ + 
               
                ~~~~~~~~~~~~~~~~~~~~~~~~~~~~~~~~~~~~~~~~~~~~~~~~$\gamma * ScoreMap_{multi}[id]$
                \State
                sorted\_evi = \textit{sortByValue}($ScoreMap_{hybrid}$)
    		\State
    		\Return sorted\_evi
                \EndFunction
    	\end{algorithmic}
        \end{center}
	
	\caption{Hybrid Ranking Algorithm.}
        \label{alg:hybrid_rank}
	\label{alg:training}
    \end{algorithm}


\subsection{Hybrid Ranking}
\noindent In FEVER, not all claims have multi-hop evidence. To maximize the overall retrieval recall, we propose a dynamic hybrid ranking algorithm to jointly rank the single-hop and multi-hop retrievals. 
Inspired by \cite{Ma2021ARS} who explored and demonstrated effective methods of combining retrieval results from dense and sparse retrievers through a simple normalization and linear combination, we demonstrate this idea also works when combining single-hop and multi-hop retrievals. 
In addition, we scale the retrieval score for each step of multi-hop retrieval through production. 
This step is important because it ensures that each episode of evidence is proportional to the other. 
A detailed implementation is presented in Algorithm~\ref{alg:hybrid_rank}.

\section{Experimental Setup}
\label{sec: Experimental_Setup}

\begin{table}[t]
\begin{center}
\resizebox{0.5\textwidth}{!}{
\begin{tabular}{l  c  c  c}
\toprule 
\textbf{Split (\#multi-hop)} & \textbf{SUPPORTS} & \textbf{REFUTES} & \textbf{NEI}\\ 
\midrule
\textbf{Train} (20,201) & 80,035 & 29,775 & 35,639  \\
\textbf{Dev} (1,960) & 6,666 & 6,666 & 6,666  \\
\textbf{Test} & 6,666 & 6,666 & 6,666  \\ 
\bottomrule
\end{tabular}}
\caption{Statistics of FEVER Dataset. The number of multi-hop claims in the blind testing set is unknown.}
\label{tab:data_partition}
\end{center}
\end{table}

\noindent This section describes the data and setup we used for our experiments.\\

\noindent\textbf{Dataset} 
Our experiments use a large-scale public fact verification dataset FEVER~\cite{thorne-etal-2018-fever}, which involves retrieving multi-hop sentence-level evidence from a large text corpus before predicting a claim’s verdict. The FEVER database comprises 185,455 annotated claims and 5,416,537 Wikipedia documents from the June 2017 Wikipedia dump. On average, each document contains 5 sentences. Annotators classify all claims as \texttt{SUPPORTS}, \texttt{REFUTES}, or \texttt{NOT ENOUGH INFO} based on single-hop and/or multi-hop sentence-level evidence.
The dataset partition is kept the same with the FEVER Shared Task~\cite{thorne-etal-2018-fever} as shown in Table~\ref{tab:data_partition}. \\

\noindent\textbf{Evaluation Metrics} As in previous work, retrieval results are compared using recall@5. Label Accuracy (LA) and FEVER score are the official evaluation metrics of the FEVER dataset. The LA metric is used to calculate the claim classification accuracy rate without considering retrieved evidence. The FEVER score checks if a complete set of golden evidence is included in the top 5 evidence retrievals in addition to the correct verdict prediction, indicating both retrieval and claim classification ability.\\

\noindent\textbf{Implementation Details}
Our best dense sentence retrievers are bi-encoder models initiated from DPR-MultiData~\cite{Karpukhin2020DensePR}. The negative examples are sampled using BM25~\cite{10.1145/3404835.3463238} and then filtered using a pre-trained attention-based sentence ranking model\footnote{\url{https://huggingface.co/cross-encoder/ ms-marco-MiniLM-L-12-v2}} at a threshold based on empirical criteria. In particular, samples with a similarity of over 0.999 are filtered out. The top two negative examples are then kept for training. Our best model is trained with a batch size of 512 and a max sequence length of 256. 
The FAISS\cite{Johnson2017BillionScaleSS} exact inner product search index (IndexFlatIP) is used to predict retrieval results that support parallel searching in GPUs.

RoBERTa-large~\cite{Liu2019RoBERTaAR} is trained for the sentence reranking module. Ten negative (\texttt{NOT ENOUGH INFO}) examples are sampled from the top 100 DSR retrievals for each claim. At inference time, we rerank the top 200 sentences retrieved from the last step (DSR).
For the final verdict prediction, we train BEVERS's~\cite{dehaven-scott-2023-bevers} claim classifier (DeBERTa-V2-XL-MNLI~\cite{he2020deberta} + XGBoost~\cite{10.1145/2939672.2939785}) with data constructed by pairing claims with M3's retrievals.
The experiments are all conducted on a machine with 8 80GB A100 GPUs. We used Huggingface Transformers~\cite{wolf-etal-2020-transformers} as the basis for our code. 

\section{Results}
\begin{table*}[t]
        \begin{center}
            \resizebox{1.0\linewidth}{!}{
        	\begin{tabular}{ll| cc cc}
        		\toprule
        	&	& \multicolumn{2}{c}{Document-level (Rec@5)} & \multicolumn{2}{c}{Sentence-level (Rec@5)} \\
        	Model Type & Model & multi-hop & Overall & multi-hop & Overall \\
        		\midrule
        	\multirow{8}{*}{Non-Iterative} &	BM25 \cite{10.1145/3404835.3463238}  & 0.252  & 0.714 &  0.385 & 0.614 \\
                & DPR-NQ \cite{Karpukhin2020DensePR} & 0.432 & 0.739 & 0.309 & 0.631 \\
                & DPR-MultiData \cite{Karpukhin2020DensePR}  & 0.452 & 0.774 & 0.320 & 0.671 \\
                & MediaWiki API + ESIM \cite{hanselowski-etal-2018-ukp} & 0.538 & -- & -- & 0.871 \\
                & MediaWiki API + BERT \cite{Soleimani-BERT-FEVER}  & -- & -- & -- & 0.884 \\
                & MediaWiki API + BM25 + T5 \cite{Jiang2021ExploringLE}  & -- & -- & -- & 0.905 \\
                & M3-DSR$_\textrm{single}$ (ours)  & 0.522 & 0.900 & 0.419 & 0.847   \\
                & M3-DSR$_\textrm{single}$+SSR$_\textrm{single}$ (ours)  & 0.633 & 0.933 & 0.572 & 0.920   \\
        	 \midrule
        	\multirow{4}{*}{Iterative} 
                & KM + Pageview + dNSMN + sNSMN + Hyperlink \cite{Nie2018CombiningFE}  & -- & 0.886 & -- & 0.868 \\
                & MediaWiki API + BigBird + Hyperlink \cite{stammbach-2021-evidence}  & 0.667 & \underline{0.945} & -- & 0.936 \\
                & TF-IDF + FSM + RoBERTa + Hyperlink \cite{dehaven-scott-2023-bevers}  & -- & -- & -- & \textbf{0.944} \\
        	&	MDR \cite{xiong2021answering}$^\dagger$ & 0.691 & -- & --&--  \\
        	&	AdMIRaL \cite{aly-vlachos-2022-natural}$^*$ & \underline{0.705} & \textbf{0.956} & --&--  \\
        	   \midrule
        	&	M3-full (ours) & \textbf{0.790} & \textbf{0.956} & \textbf{0.719} & \underline{0.940} \\
        		\bottomrule
        	\end{tabular}}
        \end{center}
	
	\caption{Retrieval performance on the FEVER dev set.  DPR-NQ and DPR-MultiData indicate the DPR model trained on the NQ dataset and the DPR-MultiData dataset, respectively by \cite{Karpukhin2020DensePR}.
DPR-MultiData dataset is a combination of multiple open-domain QA datasets consisting of NQ~\cite{Kwiatkowski2019NaturalQA}, TriviaQA~\cite{Joshi2017TriviaQAAL}, TREC \cite{Baudis2015ModelingOT}, and WQ~\cite{Berant2013SemanticPO}.
 'KM' = Keyword Matching. 'FSM' = Fussy String Matching.
  \textbf{Bold} numbers indicate best and \underline{underline} the second-best score. Iterative models are evaluated in a two-hop process, i.e., one more hop retrieval than non-iterative models. 
  }
 \label{tab:meta_retrieval}
\end{table*}

\begin{table*}
\begin{center}
    \begin{tabular}{lcc}
        \hline
        \textbf{System} & \textbf{Test LA} & \textbf{Test FEVER}\\
        \midrule
        Athene \cite{hanselowski-etal-2018-ukp} & 0.6546 & 0.6158 \\
        UNC NLP \cite{Nie2018CombiningFE} & 0.6821 & 0.6421 \\
        BERT-FEVER \cite{Soleimani-BERT-FEVER} & 0.7186 & 0.6966 \\
        KGAT \cite{Liu2019FinegrainedFV} & 0.7407 & 0.7038 \\
        LisT5 \cite{Jiang2021ExploringLE} & 0.7935 & 0.7587 \\
        BigBird-FEVER \cite{stammbach-2021-evidence} & 0.7920 & 0.7680 \\
        ProoFVer \citet{Krishna2021ProoFVerNL} & 0.7947 & 0.7682 \\
        BEVERS \cite{dehaven-scott-2023-bevers} & \underline{0.8035} & \textbf{0.7786} \\
        \midrule
        M3-FEVER (ours) & \textbf{0.8054} & \underline{0.7743} \\
        \midrule
    \end{tabular}
\end{center}

\caption{
Full system comparison for label accuracy (LA) and FEVER score on the blind FEVER test set. \textbf{Bold} numbers indicate the best and \underline{underline} the second-best score.
	}
\label{tab:fever_test_results}
\end{table*}

Following the workflow of our multi-hop retriever (M3), we report the evaluation of the five major components in M3 sequentially: (1) Single-hop Dense Sentence Retrieval, (2) Single-hop Sentence Reranking, (3) Multi-hop Dense Sentence Retrieval, (4) Multi-hop Sentence Reranking, and (5)Dynamic Hybrid Ranking.
\vspace{-2pt}
\subsection{Evidence Retrieval}
\noindent A summary of the major retrieval evaluation results can be found in Table~\ref{tab:meta_retrieval}.
We evaluate the multi-hop and the overall retrieval performance at two levels of granularity, namely the document and sentence levels. Results include non-iterative retrievers, covering sparse retrieval (BM25), dense passage retrieval (DPR), and multi-stage retrieval methods, i.e., document retrieval + sentence selection (reranking). MediaWiki API\footnote{\url{https://www.mediawiki.org/wiki/API:Main_page}} is one of the most commonly used methods for document retrieval. It searches through the titles of all Wikipedia articles for entries that match the entity mentions found in the claim. \citet{Jiang2021ExploringLE} combine the BM25 and WikiMedia API results by going through the two ranked lists of documents alternately, skipping duplicates, and keeping the top $k$ unique documents. Different attention-based sentence selection (reranking) methods are used, such as Enhanced Sequential Inference Model (ESIM)~\cite{hanselowski-etal-2018-ukp}, BERT~\cite{Soleimani-BERT-FEVER}, and T5~\cite{Jiang2021ExploringLE}. 

We further compare M3 against state-of-the-art iterative retrieval approaches, including MDR, and AdMIRaL which have been introduced in Section~\ref{sec:related-works}. We also compare with those methods that rely on hyperlinks for multi-hop retrieval~\citep{Nie2018CombiningFE, stammbach-2021-evidence, dehaven-scott-2023-bevers}. 
Specifically, 
\cite{stammbach-2021-evidence} used MediaWiki API for single-hop document retrieval, while \cite{Nie2018CombiningFE} and \cite{dehaven-scott-2023-bevers} used complex combined methods, i.e., \cite{Nie2018CombiningFE} used keyword matching + pageview frequency +
neural network-based document reranker (dNSMN);
\cite{dehaven-scott-2023-bevers} combined results from TF-IDF on titles, TF-IDF on content, and fuzzy string matching on titles queried by entities extracted from claims. Different neural network-based pairwise scoring models are used for sentence reranking, i.e., sNSMN~\cite{Nie2018CombiningFE}, BigBird~\cite{stammbach-2021-evidence}, and Roberta-large~\cite{dehaven-scott-2023-bevers}.

As M3 supports only sentence-level retrieval, when calculating document-level recall@5, we compare the golden document IDs with those of our top 5 sentence retrievals. It's important to note that this is a harder setting for us, since on average, only 2.87 document IDs are included in our top 5 sentence retrievals. Despite this setting, M3 still achieves the highest multi-hop and overall retrieval recall for documents, and the highest multi-hop retrieval recall for sentences, only trailing \cite{dehaven-scott-2023-bevers} on retrieval recall for overall sentences. 

\subsection{End-to-end Fact Verification}
\noindent Furthermore, we test our fact verification system M3-FEVER in an end-to-end manner on the FEVER blind test set. As shown in Table~\ref{tab:fever_test_results}, M3-FEVER achieves the highest LA score and the second-best FEVER score\footnote{The FEVER Official Leaderboard: \url{https://codalab.lisn.upsaclay.fr/competitions/7308}}.

\section{Analysis}\label{sec: analysis}
\noindent This section examines the effects of different major design decisions made in our M3 system.

\subsection{Effect of multi-task learning}
\begin{figure}[t]
    \centering
    \includegraphics[width=\linewidth]{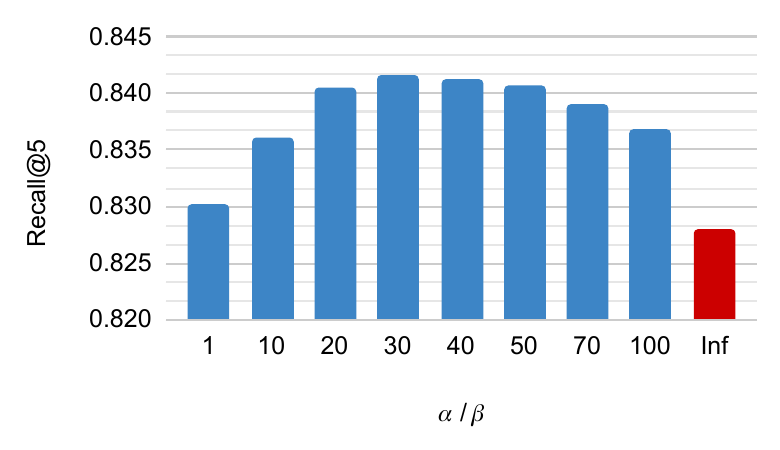}
    \caption{M3-DSR$_\textrm{single}$ top-5 retrieval recall with different ratios of multi-task learning loss weights, where $\alpha$ and $\beta$ represent the weight of contrastive loss and claim classification loss, respectively. 'Inf' indicates only using the contrastive objective during training, i.e., single-task learning.}
    \label{fig:recall_multi-task_loss_ratio}
\end{figure}

\noindent 
We explore what ratio of multi-task learning loss weights, i.e.$\alpha / \beta$ in Equation~\ref{eq:joint_loss}, is optimal for dense sentence retrieval.
Figure~\ref{fig:recall_multi-task_loss_ratio} illustrates the top-5 sentence retrieval recall of M3-DSR$_\textrm{single}$ with respect to different $\alpha / \beta$, measured on the FEVER dev set, where $\alpha$ and $\beta$ represent the weight of contrastive loss and claim classification loss, respectively.
As is shown,  when $\alpha / \beta=30$ gives the highest retrieval recall. It outperforms the sole contrastive object learning by 1.65\%. This suggests that our multi-task learning framework is effective in learning higher-quality dense sentence representations.

\subsection{Effect of mixed-objective learning}
\begin{figure}[t]
    \centering
    \includegraphics[width=\linewidth]{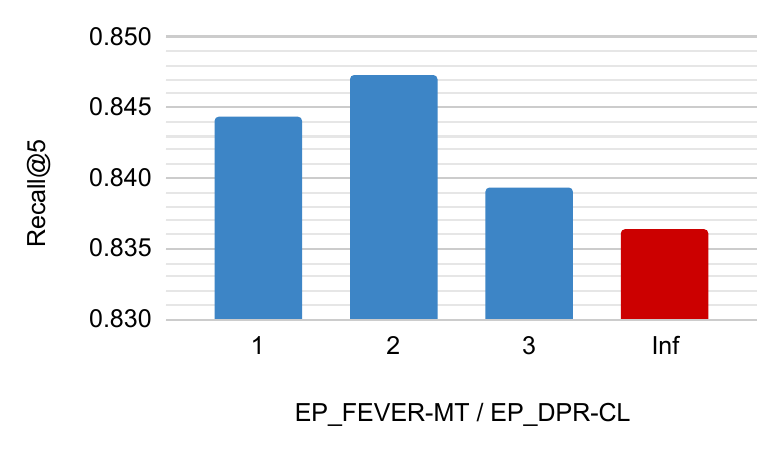}
    \caption{M3-DSR$_\textrm{single}$ top-5 retrieval recall with different ratios of mixed-objective training epochs.
    'Inf' indicates that only the FEVER dataset is used for training with the multitask learning objective.}
    \label{fig:recall_mix_objective_ratio}
\end{figure}

\noindent
We trained M3-DSR on two datasets with different objectives jointly using mixed-objective learning. Specifically, we train M3-DSR alternatively on the DPR-MultiData dataset with the contrastive objective and on FEVER with the multi-task objective.
We test different ratios of training epochs for the two datasets when doing mixed-objective learning in Figure~\ref{fig:recall_mix_objective_ratio} and observe that when EP$_\textrm{FEVER-MT}$ / EP$_\textrm{DPR-CL}$ = 2 (i.e., training on the FEVER dataset with the multi-task object for two epochs after every epoch of training on the DPR-MultiData dataset with the contrastive object) gives the best performance. This indicates that our mixed-objective learning framework is effective at learning dense sentence representations with higher quality.

\subsection{Effect of hybrid-ranking algorithm}
We compare our hybrid-ranking method with two different types of merging algorithms: 1) Threshold: jointly rank multi-hop retrievals whose scores are larger than a threshold with the single-hop retrievals. 2)Scale~\citep{stammbach-2021-evidence, dehaven-scott-2023-bevers}: re-scale the multi-hop retrievals by a factor before jointly ranking them together with the single-hop retrievals.
Table~\ref{tab:result_hybrid_ranking} demonstrates that our hybrid-ranking algorithm outperforms the baseline algorithms by a large margin.

\begin{table}[t]
\centering
\begin{tabular}{lc}
\hline
\textbf{Method} & \textbf{Recall@5}\\
\midrule
Threshold & 0.925\\
Scale & 0.931\\
Hybrid Ranking & \textbf{0.940}\\
\bottomrule
\end{tabular}
\caption{
Ablation of the hybrid ranking algorithm over the FEVER's dev set. All hyperparameters are tuned through grid search over our best SRR$_\textrm{multi}$'s results.
}
\label{tab:result_hybrid_ranking}
\end{table}
 
\subsection{Effect of negative sampling}
\noindent Due to the difficulty of exhaustively annotating all positive examples given a query, false negatives are common in large-scale retrieval datasets. Figure~\ref{fig:negative_sampling} demonstrated false negative examples in the FEVER dataset. When using traditional sampling methods such as BM25 to sample negative examples for contrastive learning, we find it difficult to avoid false negatives. By applying an empirical threshold to an off-the-shelf attention-based ranking model, we can eliminate more false negatives from training data, thereby further improving M3-DSR$_\textrm{single}$'s recall by 5.6\%.


\begin{figure}[h]
\begin{center}
\scalebox{1.0}{
\begin{tabular}{@{}p{21em}}
\toprule 
\textbf{Claim: Romelu Lukaku plays in the Premier League for Everton.} \\
\midrule
\textbf{Single-hop evidence annotations:}\\
\vspace{1pt}
1. (Title: Romelu Lukaku) Romelu Menama Lukaku ( born 13 May 1993 ) is a Belgian professional footballer who plays as a striker for Premier League club Everton and the Belgium national team.\\
\vspace{0.2pt}
2. (Title: Romelu Lukaku) He did not appear regularly in his first season there, and spent the following two seasons on loan at West Bromwich Albion and Everton respectively, signing permanently for the latter for a club record \# 28 million in 2014.\\

\midrule
\textbf{Top-2 sampled negatives by BM25:}\\
\vspace{1pt}
1. (Title: Lukaku) \textcolor{red}{\textit{Romelu Lukaku ( born 1993 ), Belgian footballer, who currently plays for Everton.}}\\
\vspace{0.2pt}
2. (Title: Roger Lukaku) He is the father of footballers Romelu Lukaku and Jordan Lukaku. \\
\midrule
\textbf{Verdict:} Supported\\
\bottomrule
\end{tabular}}
\end{center}
\caption{
An example of a false negative sampled by BM25 from the FEVER is highlighted in red.
}
\label{fig:negative_sampling}
\end{figure}

\section{Conclusion}
In this paper, we introduce M3, an advanced recursive multi-hop dense sentence retrieval system designed for fact verification. M3 achieves top-tier performance in multi-hop retrieval on the FEVER dataset. We propose a novel method for learning dense sentence representations, which is based on multi-task learning and mixed-objective learning. This approach addresses challenges faced by current dense retrieval methods that rely on contrastive learning. Furthermore, we present an efficient heuristic hybrid ranking algorithm that combines single-hop and multi-hop sentence evidence, resulting in significant improvements over previous methods. Lastly, we develop an end-to-end multi-hop fact verification system built upon M3, which also attains state-of-the-art performance on the FEVER dataset.

\section{Acknowledgements}
Our sincere thanks go out to the anonymous reviewers who took the time to provide constructive comments. 

\section{Ethics Statement}
We expect our retrieval system to be utilized in fact-checking applications. It's important to clarify that our system doesn't assess the accuracy of real-world statements; it solely relies on Wikipedia as its source of evidence, as our entire testing environment is limited to this dataset. While Wikipedia is a valuable collaborative resource, it's not immune to errors and inaccuracies, just like any other encyclopedia or knowledge base. Therefore, we advise users against using our retrieval system to make definitive claims about the accuracy of the statements being verified. In other words, it should not be employed as a tool to determine the absolute truth of claims – in other words, avoid using it to declare statements as true or false.


\section{References}\label{sec:reference}
\vspace{-20pt}

\bibliographystyle{lrec-coling2024-natbib}
\bibliography{custom}

\end{document}